\documentclass[12pt]{article}
\usepackage{amsthm,amssymb,amsmath}
\newtheorem{Th}{Theorem}
\newtheorem{pro}{Proposition}
\newtheorem{defi}{Definition}
\newtheorem{lemma}{Lemma}
\newtheorem{cor}{Corollary}

\newcommand{\dif}{\operatorname{d}}
\newcommand{\I}{\operatorname{i}}
\newcommand{\re}{\operatorname{Re}}

 \newcommand{\res}{\operatorname{res}}
\newcommand{\GL}{\operatorname{GL}}

\newcommand{\bt}{\boldsymbol t}

\newcommand{\C}{{\mathbb C}}
\newcommand{\R}{{\mathbb R}}

\newcommand{\N}{{\mathbb N}}
\newcommand{\id}{{\mathbb I}}
\newcommand{\g}{\mathfrak g}

\newcommand{\D}{{\partial}}

\begin{document}

\title{Vectorial Darboux Transformations for\\ the Kadomtsev-Petviashvili 
Hierarchy}

\author{Q. P. Liu\thanks{On leave of absence from
Beijing Graduate School, CUMT, Beijing 100083, China}
\thanks{Supported by {\em Beca para estancias temporales
de doctores y tecn\'ologos extranjeros en
Espa\~na: SB95-A01722297}}
   $\,$ and M. Ma\~nas\thanks{Partially supported by CICYT:
 proyecto PB95--0401}\\
 Departamento de F\'\i sica Te\'orica,
\\ Universidad Complutense,\\
E28040-Madrid, Spain.}

\date{}

\maketitle

\begin{abstract}
We consider the vectorial approach to the binary Darboux
transformations for the Kadomtsev-Petviashvili hierarchy in its
Zakharov-Shabat formulation. We obtain explicit formulae for the
Darboux transformed potentials in terms of Grammian type
determinants. We also  study the $n$-th Gel'fand-Dickey hierarchy
introducing spectral operators
and obtaining similar results. We
reduce the above mentioned results to the
 Kadomtsev-Petviashvili I and II real
forms, obtaining corresponding vectorial Darboux transformations.
In particular for the Kadomtsev-Petviashvili I hierarchy we get the
line soliton, the lump solution and the Johnson-Thompson lump, and
the corresponding determinant formulae for the non-linear superposition
of several of them. For Kadomtsev-Petviashvili II apart from the line solitons we 
get
singular rational solutions with its singularity set describing the
motion of strings in the plane. We also consider the I and II real
forms for the Gel'fand-Dickey hierarchies obtaining the vectorial
Darboux transformation in both cases.
\end{abstract}
\newpage

\section{Introduction}
The Kadomtsev-Petviashvili (KP) hierarchy is a corner-stone in the
theory of integrable systems, and its relevance in Mathematics and
Physics is well established, see \cite{ac,kp,k,nm} and references
therein. The KP hierarchy is a set of commuting flows that receives
its name from one of the equations contained in it: the KP
equation. The KP equation was found for the first time in studies
on plasma physics \cite{kpe} and it is also relevant in the theory of
surface water waves \cite{ase}. Moreover, the equation is the
compatibility condition of two differential linear operators
\cite{zs,dy} and one of them is precisely
the non-stationary Schr\"{o}dinger
operator. There
are two relevant real reductions, known as KPI and KPII, and the
inverse spectral transform for KPI was performed in \cite{istkpi,fa}
and for KPII in \cite{istkpii}, for a more analytical point of view see
 \cite{fs}. For more details, such as type of solutions,
 hamiltonian
structures, symmetry groups, etc.,  on the equation we refer the
reader to \cite{ac} and references therein.
Apart form its relevance in Physics, the KP
equation is  important in Mathematics as well. In particular,
let us mention its deep connection with the theory of Riemman surfaces
and theta functions, and the Novikov conjecture on the Schottky problem, see
\cite{sch,nm,kp,shi}.
Let us mention that recently Fokas and Zakharov \cite{fz} have extended the 
inverse
scattering method and applied it, in particular, to the $N$-wave resonant 
interaction,
Davey-Stewartson I an KPI equations.
For this aim they consider a non-local Riemann-Hilbert problem.  Futher 
comparasions
between this method and the results of our paper is a work in progress.

This hierarchy contains many relevant integrable equations such as
the Korteweg-de Vries (KdV) and Boussinesq, which are particular
examples of the Gel'fand-Dickey (GD) hierarchies \cite{gd}. The GD
hierarchies appear also as particular reductions of the KP
hierarchy \cite{kp}. The relevance of the KdV and Boussinesq
equations in fluid dynamics \cite{ac} has been known for long time,
but  recently  the KdV hierarchy has been found to play an
important role in a non-perturbative analysis of 2 dimensional
quantum gravity, and thus the consequent relevance in this field of
the KP equation, see \cite{hmm}. From here it steams the relevance
of the KdV and KP hierarchies in the analysis of the moduli space
for the intersection theory of complex curves \cite{wit}.

At the end of the last century Darboux studied certain
transformation of the Schr\"{o}dinger equation that provided new
potentials and wave functions at the same time \cite{darboux}. This
method has been very much extended and applied to a large number of
integrable systems \cite{ms}. The Darboux transformation for the KP
equation was given by Matveev in \cite{matveev},
and the whole hierarchy, initially considered in \cite{matveev},
  was recently considered, in its Lax form, by Oevel and
Rogers \cite{or} ---see also \cite{oevel,chau}.
The binary extension of the Darboux transformation for the KdV equation
was first introduced in \cite{levi,matveev2}, but as was observed in
\cite{athorne} it is just an adequate composition of standard Darboux
transformations.

One can iterate the binary Darboux transformations to obtain
formulae of Gramm type for the KP hierarchy \cite{os} (for the KP
equation see \cite{ms,athorne}). In this paper we observe that in
fact there is an alternative way to obtain these results. The
simple and basic idea is to replace the scalar character of the
initial wave functions by a vectorial one. We will show that this
replacement gives explicit compact expressions in terms of Gramm
type determinants for all the potentials appeared in the
zero-curvature representation of the KP hierarchy, avoiding in this
manner the tedious iteration. Also we study how this technique
reduces to the GD hierarchy. Finally, we show that one of the main
advantages of the vectorial approach appears when real reduction is
concerned. In fact, we give vectorial Darboux transformations for
both KPI and KPII hierarchies, the real reductions of the KP
hierarchy. Regarding the KPI hierarchy we obtain the following
novel results:
\begin{enumerate}
\item Departing from any solution we obtain vectorial Darboux transformations,
expressed in terms of Gramm type determinants, that preserve the
singularity structure of the solution; i. e., if the departing
solution is singularity free so is the Darboux transformed one.
\item Next, we dress the vacuum solution using heat quasipolynomials.
 In doing so we obtain compact
determinant formulae that contain both the $N$ line soliton
solutions and the lump solutions, and also  mixture of them. In
particular, determinant expressions are given for non-singular
rational solutions, containing not only the standard lump solutions
\cite{mz,as} but also the more exotic lump solutions presented for
the first time in \cite{jt} and recently studied  by Ablowitz and
Villarroel in \cite{av}.
\item The $\tau$ functions for these solutions, which are
strictly positive as we shall prove, are given as determinant of
 Galilean shifted heat polynomials. For the scalar case we also provide an
 expression, a sum of square modulus of complex polynomials,
which explicitly shows the positivity
 of the $\tau$-function. This expression  contains in particular
 the 1-lump \cite{mz,as}, the
 Johnson-Thompson one \cite{jt} and the more involved solutions
as those constructed with
 higher degree heat polynomials.
\end{enumerate}

Let us mention that the use of heat polynomials
for the construction of  rational solutions of KPI was
first propose in \cite{matveev3}, see also \cite{peli}.

For the KPII reduction we get a general vectorial Darboux
transformation and as an example we find the line solitons and
rational singular solutions. For the KPII equation these
singularities are algebraic curves in the plane which evolve in
time.

Finally we consider the two real reductions for the GD hierarchy,
that is GDI and GDII hierarchies. We present for both cases
different schemes providing vectorial Darboux transformations. We
remark that the KdV hierarchy (the 1st GD hierarchy) is a very
special case for which both KdVI and KdVII hierarchies coincide in
the real KdV hierarchy, thus for the real KdV hierarchy one has at
hand two unequivalent vectorial Darboux transformations. However,
for the other GD hierarchies the two reductions are different as
 the Boussinesq (2nd GD hierarchy) example shows: one has two
Boussinesq equations (however only the Boussinesq I has
known physical applications); but for both we have vectorial
Darboux transformations.

We need to emphasize that the vectorial Darboux transformation
presented has at least two advantages: avoiding the iteration
process and a simple implementation of the real reduction problem.
Also, as any Darboux transformation it not only provides new
potentials, it also gives new wave functions. The relevant feature
of this is that the first linear equation is just the
non-stationary Schr\"{o}dinger equation. From here it steams  the
relevance of the results of this paper not only in relation with
the KP equation but also with the non-stationary Schr\"{o}dinger
equation.

The layout of the paper is as follows: in \S 2 we recall
 the Zakharov-Shabat
representation of the KP hierarchy and construct the corresponding
vectorial Darboux transformation. We explicitly present expressions
 for all the Darboux
transformed potential functions. Next, in \S 3, we reduce these
results to the GD case. \S 4 is devoted to the real reductions KPI
and KPII. The real reductions GDI and GDII are considered
in \S 5.

\section{Vectorial Darboux transformations for the KP hierarchy}

In this section we study the vectorial extension of the binary Darboux
transformation for the KP hierarchy. This hierarchy has several
representations,
and as for the vectorial Darboux transformation
is concerned the most appropriate one is the
Zakharov-Shabat or zero-curvature
representation. Let us point out that the more usual one is
the Lax representation,
however it is a well-known fact that  these two are actually equivalent
\cite{takebe}.

We shall call KP hierarchy to the infinite set of compatible
equations ---Zakharov-Shabat equations or
 zero-curvature conditions---
\begin{equation}\label{kp}
\partial_n (B_m)-\partial_m(B_n)+[B_m,B_n]=0,\;\;\; m,n\in\N,
\end{equation}
where $B_m$, $m\in\N$, are polynomials in
$\partial:=\partial/\partial x$; i. e., differential operators in the
independent variable $x$, of the following form:
\begin{equation}\label{B}
B_m=\partial^m+u_{m,m-2}\D^{m-2}+\dotsb+u_{m,0},
\end{equation}
where the dependent variables $u_{i,j}$ are functions of
$t_1:=x,t_2,t_3\dotsc$. Here we are using the notation
$\partial_m=\D/\D t_m$. For example, the first non trivial case
corresponds to $m=2,n=3$ for which $u_{2,0}=:u$, $\D u_{3,1}=3/2\D u$, so 
that
we can take  $u_{3,1}=3/2 u+c$, $c$ an arbitrary constant and
$\D(u_{3,0})=3/4(\D_2(u)+\D^2(u))$ and the well known KP equation
appears:
\[
\D\Big(\D_3u-\frac{1}{4}\D^3u-\Big(\frac{3}{2} u+c\Big) \D 
u\Big)=\frac{3}{4}\D_2^2u.
\]


Observe that the KP hierarchy (\ref{kp}) is the set of
compatibility conditions for the following linear system
\begin{equation}\label{linear}
\D_m b=B_m(b), \;\;\; m\in\N,
\end{equation}
where $b$ is usually assumed to be a scalar function of
 $\bt:=\{t_m\}_{m\in\N}$. However, this also holds
when $b$ takes values in an arbitrary linear space $V$.
Note  that the KP hierarchy is as well
 the compatibility condition for the adjoint
linear system
\begin{equation}\label{adlinear}
\D_m \beta=-\tilde B_m(\beta), \;\;\; m\in\N,
\end{equation}
where
\[
\tilde B_m:=(-1)^m\partial^m+(-1)^{m-2}\D^{m-2}u_{m,m-2}+
(-1)^{m-3}\D^{m-3}u_{m,m-3}+\dotsb+u_{m,0}.
\]
As before,  $\beta$ is usually a scalar function of $\bt$
and once again we can give it a vector character,
in particular we take it
as a linear functional in $V$; i. e., as an element in the
dual space $V^*$ of $V$.

Fixed $\beta$, we now introduce a potential operator
$\Omega(\bt)\in\GL(V)$, depending on an arbitrary wave function
$b$,
 through the following compatible equations ---this compatibility can be proven
 as in the scalar case \cite{oevel}:
\begin{equation}\label{potential}
\D_n(\Omega)=-\res\big(\D^{-1}b\otimes \tilde B_n\beta\D^{-1}\big),
\end{equation}
where the residue (res) of a pseudo-differential operator is the coefficient
of $\D^{-1}$.

Associated with $b$, $\beta$ and $\Omega$ we now define fundamental
scalar functions which will be needed
in the construction of the Darboux transformation.

\begin{defi}
Given a solution of the KP hierarchy $\{u_{n,m}\}$, a vector wave
function $b$ solving (\ref{linear}) and  an adjoint wave function
$\beta$ solving (\ref{adlinear}), we introduce
\begin{equation}\label{f}
f_\ell:=\langle \beta,\Omega^{-1} b^{(\ell)}\rangle.
\end{equation}
\end{defi}

In this paper we understand $f^{(\ell)}=\D^\ell(f)$.
 Related to these quantities one has
\[
g_\ell:=\langle
\beta,(\Omega^{-1} b)^{(\ell)}\rangle,
\]
 that can be
expressed in terms of the $f$'s. To this end, we introduce
\begin{equation}\label{calg}
{\cal G}_{i,j}:=
\sum_{k+\ell=j-1}\binom{i-k-1}{\ell}f_k^{(\ell)},\;\;\; 0<j\leq i,
\quad {\cal G}_{i,0}=1.
\end{equation}

\begin{pro}
One can express $g_\ell$ as a differential polynomial in the $f$'s as
follows:
\begin{equation}\label{g}
\begin{aligned}
g_\ell=&f_\ell-f_0\;{\cal
G}_{\ell,\ell}\\
&-\sum_{k=1}^{\ell-1}\Bigg[\Big({\cal
G}_{\ell,k}+\sum_{m=1}^{\ell-2}(-1)^m
\sum_{\ell>k_1>\dotsb>k_m>k}{\cal
G}_{\ell,k_1}{\cal G}_{k_1,k_2}\dotsb{\cal
G}_{k_m,k}\Big)(f_k-f_0\;{\cal G}_{k,k})\Bigg],
\end{aligned}
\end{equation}
with ${\cal G}_{i,j}$ as in (\ref{calg}).
\end{pro}
\begin{proof}
It is convenient to
 compute $\Omega^{-1}b^{(k)}$:
\[
\Omega^{-1}\D^k(b)=\Omega^{-1}\D^k(\Omega \hat{b})=
(\Omega^{-1}\D\Omega)^k(\hat b)=(\D+\Omega^{-1}\D(\Omega))^k(\hat{b}),
\]
where
 $\hat{b}=\Omega^{-1}b$.
 Equation (\ref{potential})  implies in particular  that
\begin{equation}\label{potenx}
\D(\Omega)=b\otimes\beta,
\end{equation}
and so
\[
\Omega^{-1}b^{(k)}=(\D+\hat{b}\otimes\beta)^k(\hat{b}).
\]

Therefore, we have
\begin{equation}\label{id1}
\Omega^{-1}b^{(k)}=\hat{b}^{(k)}+\sum_{j=1}^{k}{A}_{k,j}
\hat{b}^{(k-j)}.
\end{equation}
Using this formula one arrives at some recursion relations
among the coefficients $A_{m,j}$, which are uniquely
satisfied by the choice $A_{i,j}={\cal G}_{i,j}$ of
 (\ref{calg}). Contracting this expression with the linear
functional $\beta$ one obtains the following
inhomogeneous linear system
\[
\begin{pmatrix}
1&0&0&\dots&0\\ {\cal G}_{2,1}&1&0&\dots&0\\
{\cal G}_{3,2}&{\cal G}_{3,1}&1&\ddots&\vdots\\
\vdots&\vdots&\ddots&\ddots&0\\
{\cal G}_{m,m-1}&{\cal G}_{m,m-2}&\dots&{\cal G}_{m,1}&1
\end{pmatrix}
\begin{pmatrix}g_1\\ g_2 \\ g_3
\\ \vdots \\ g_m\end{pmatrix}
=\begin{pmatrix} f_1-f_0\;{\cal G}_{1,1}\\  f_2-f_0\;{\cal G}_{2,2}
 \\  f_3-f_0\;{\cal G}_{3,3}
\\ \vdots \\ f_m-f_0\;{\cal G}_{m,m}\end{pmatrix}.
\]
Solving the above linear system we get the desired result.
\end{proof}

Observe that the functions $f$'s are expressed in terms of the
contraction of $\beta$, $\Omega^{-1}$ and $x$-derivatives of $b$.
>From (\ref{potenx}), if $\Omega$ has a determinant, one deduces that
\[
f_0=\D\ln\det\Omega,
\]
and, in the finite-dimensional case,  the Cramer's rule implies
\[
f_j=\sum_k \beta_k \frac{\det \Omega^{[k,j]}}{\det\Omega},
\]
where $\Omega^{[k,j]}$ is obtained from the corresponding matrix of
$\Omega$ by replacing the $k$-th column by the column given by
$b^{(j)}$.

The operator $\Omega$ resembles a Gramm matrix.
 In fact, from (\ref{potenx})   we have
\[
\Omega=C+\D^{-1}(b\otimes\beta),
\]
with $\D^{-1}$  a primitive of $\D$ and  $C(\bt)$ a linear operator
 with $\D
(C)=0$, for vanishing initial data  $x=-\infty$ the operator $C$ could
be chosen as a constant operator.

To proceed further  we introduce some convenient notation:
\begin{defi}
\begin{align*}
h_{n,k}:=&\sum_{\ell =k}^{n}\sum_{i=0}^{\ell-k}(-1)^{i+\ell}
\binom{\ell-k}{i}\big(u_{n,\ell}\;g_{i}\big)^{(\ell-k-i)},\;\; 0\leq k
\leq n,\\
 h_{n,n+1}:=&0.
\end{align*}

\end{defi}
With this at hand we can show:
\begin{lemma} Given a solution of the KP hierarchy $\{u_{n,m}\}$,
a function $b$ solving (\ref{linear}) and   an adjoint wave
function $\beta$ solving (\ref{adlinear}),  then the functions
$\hat{b}:=\Omega^{-1}b$ and $\hat \beta:=\beta\Omega^{-1}$ satisfy
\begin{align*}
\D_n \hat{b}&=\hat B_n(\hat{b}),\\
\D_n\hat{\beta}&=-\Tilde{\Hat{B}}_n(\hat\beta),
\end{align*}
where
\begin{equation}\label{hatu}
\Hat B_n=\sum_{m=0}^n \hat u_{n,m} \D^m,\;\;\;
\hat u_{n,m}:=\sum_{j=m}^{n}{\cal G}_{j,m}\big(u_{n,j}+(-1)^jh_{n,j+1}\big).
\end{equation}
\end{lemma}
\begin{proof}From $b=\Omega\hat b$ it follows that
\begin{equation}\label{eq}
\D_n\hat b=\Omega^{-1}\left(B_n(\Omega\hat b)+\res (\D^{-1}b\otimes
\tilde B_n\beta
\D^{-1})\hat b\right),
\end{equation}
and  using  identity (\ref{id1}) we arrive to
\[
\Omega^{-1}B_n(\Omega\hat b)=\sum_{m=0}^{n}\left(\sum_{j=m}^{n}
u_{n,j}{\cal G}_{j,j-m}\right)\hat b^{(m)},
\]
and
\[
\Omega^{-1}\res(\D^{-1}b\otimes\tilde B_n\beta \D^{-1})\hat b=
\sum_{m=0}^{n-1}\left(\sum_{j=m}^{n-1}h_{n,j+1}{\cal G}_{j,j-m}
\right)\hat b^{(m)}.
\]
The substitution of the last two formulae into (\ref{eq}) proves
the lemma.
\end{proof}

Observe that the differential operator  has the same form as
$B_n$; i. e., the highest coefficient is 1 and the next to highest
vanishes. The first non-trivial coefficient in $\hat B_n$ is
\[
\hat u_{n,n-2}=u_{n,n-2}+n \D (f_0).
\]

We are ready to prove the following result
\begin{Th}
The functions
\begin{equation*}
\hat u_{n,m}:=\sum_{j=m}^{n}{\cal G}_{j,m}\big(u_{n,j}+(-1)^jh_{n,j+1}\big),
\end{equation*}
with
\begin{align*}
{\cal G}_{i,j}&:=
\sum_{k+\ell=j-1}\binom{i-k-1}{\ell}f_k^{(\ell)},\;\;\; 0<j\leq i,\quad
{\cal G}_{i,0}=1,\\ h_{n,k}&:=
\begin{cases}\sum_{\ell =k}^{n}\sum_{i=0}^{\ell-k}(-1)^{i+\ell}
\binom{\ell-k}{i}\big(u_{n,\ell}\;g_{i}\big)^{(\ell-k-i)},& 0\leq k\leq n,\\
 0,& k=n+1,
\end{cases}\\
f_\ell&:=\langle \beta,\Omega^{-1} b^{(\ell)}\rangle,\\
g_\ell&:=\langle
\beta,(\Omega^{-1} b)^{(\ell)}\rangle,
\end{align*}
 solve
the KP hierarchy.
\end{Th}
\begin{proof} Given an arbitrary solution $\psi(\bt)\in V$ of
(\ref{linear}) we take the following solution of (\ref{linear})
\[
\underline{b}:=\left(\begin{array}{c}\psi\\ b\end{array}\right)\in V\oplus V,
\]
and a solution of (\ref{adlinear}) as
\[
\underline{\beta}:=(0,\beta)\in V^*\oplus V^*,
\]
then  a potential operator $\underline{\Omega}$, defined by
$\underline{b}$ and $\underline{\beta}$ as prescribed by
(\ref{potential}), is given by
\[
\underline{\Omega}=\left(\begin{array}{cc} 1&\check\Omega \\
0& \Omega\end{array}\right),
\]
where $\check\Omega$ is defined by (\ref{potential}) replacing $b$
by $\psi$. Now, the transformed wave function $\Hat{\underline{b}}$
is
\[
\Hat{\underline{b}}=\underline{\Omega}^{-1}\left(\begin{array}{c}\psi\\ 
b\end{array}\right)=
\left(\begin{array}{c}\psi-\check\Omega \Omega^{-1}b\\
\Omega^{-1}b\end{array}\right),
\]
while
\[
f_j=\langle\underline{\beta},\underline{\Omega}^{-1}\underline{b}^{(j)}\rangle=
\langle\beta,\Omega^{-1}b^{(j)}\rangle
\]
is not changed. So, we can use the above lemma, that tells us that
$\underline{b}$ does satisfy (\ref{linear}) with the coefficients
$\hat u_{n,m}$, and deduce that so does
$\psi-\Check\Omega\Omega^{-1}b$ for an arbitrary departing wave
function $\psi$. Therefore, following standard arguments \cite{ms},
we conclude that $\hat B_n$ must satisfy (\ref{kp}).

\end{proof}

>From the proof of the theorem we can extract the following:

\begin{cor}
Given an arbitrary initial wave function $\psi$ the vectorial
Darboux transformation associated with the data $b,\beta$ and
$\Omega$ as in the previous theorem provides a new wave function
$\Hat\psi$ satisfying the new linear system. Namely:
\[
\Hat\psi:=\psi-\Check\Omega\Omega^{-1}b,
\]
with $\check\Omega$  defined by (\ref{potential}) replacing $b$ by
$\psi$.
\end{cor}

\section{Reduction to the Gel'fand-Dickey hierarchy}

It follows from the equivalence between the Lax and Zakharov-Shabat
representations of the KP hierarchy that the
$n$-th GD reduction appears when the solutions
$\{u_{i,j}\}$ do not depend on $t_n$. In fact, if this is
the case,
they do not depend on the times $\{t_{nm}\}_{m\in\N}$. The other flows
describes isospectral transformations of the differential operator
\[
{\cal L}:=B_n=\D^n+u_{n-2}\D^{n-2}+\dotsb+u_0,
\]
where we use the notation
\[
u_j=u_{n,j},
\]
in terms of which all the $u_{m,i}$ are expressed as differential
polynomials.

Regarding the vectorial Darboux transformation the point is to
choose $b$, $\beta$ and $\Omega$ in an appropriate way so that the
GD reduction is preserved. That is, given a solution $\{ u_{i,j}\}$
of the KP hierarchy not depending on $t_n$ choose those Darboux
transformations for which the Darboux transformed solution, given
by $\{\hat u_{i,j}\}$  as in (\ref{hatu}),  does not depend on
$t_n$. Because the new solution $\hat u$ is constructed in terms of
the $u$ and the functions $f$'s, defined by (\ref{f}), this aim
will be achieved if the latter functions do actually not depend on
$t_n$.

The pertinent choice in this case, as we shall proof later, for $b$
and $\beta$  is given by
\begin{equation}\label{lax}
\D_nb={\cal L}(b)=L b, \;\; \D_n\beta=-{\cal L}^*(\beta)=-\beta\Lambda,
\end{equation}
where $L$ and $\Lambda$ are arbitrary linear operators in $V$.
These two linear operators could be considered as noncommutative
extensions of spectral parameters. Observe also that in general
they can not be got from an standard iteration of the scalar binary
Darboux transformation: to get non-diagonalizable spectral
operators one needs to perform, apart from the iteration, suitable
coalescence of  eigen-values.

When $b$ and $\beta$ are chosen in this particular form one has
\begin{pro}
If $\{u_{i,j}\}$ is a solution of the KP hierarchy  not depending on
$t_n$, and $b$ and $\beta$ are chosen as solutions of (\ref{lax}),
 the operator $\Omega$ satisfies
\[
\partial_m(\partial_n\Omega-L\Omega+\Omega\Lambda)=0,\;\;\; m\in\N.
\]
\end{pro}
\begin{proof}
>From the compatibility condition $\D_m\D_n\Omega=\D_n\D_m\Omega$
and (\ref{potential}) one easily derives
\begin{align*}
\D_m\D_n\Omega=&-\D_n\left(\res(\partial^{-1}b\otimes
\tilde B_{m}\beta\partial^{-1})\right)\\
=&
-\res\Big(\partial^{-1}\Big[(\D_nb)\otimes \tilde B_{m}\beta+
b\otimes (\D_n\tilde B_{m})\beta+b\otimes
\tilde B_{m}(\D_n\beta)\Big]\partial^{-1}\Big).
\end{align*}
As we are dealing with a solution that reduces to the $n$-th GD
hierarchy we have $\D_n(B_m)=0$, and because (\ref{lax}) holds we find
that
 \begin{align*}
\D_m\D_n\Omega&=-L\res(\partial^{-1}b\otimes \tilde B_{m}\beta\partial^{-1})
+\res(\partial^{-1} b\otimes
\tilde B_{m}\beta\partial^{-1})\Lambda\\
&=\D_m(L\Omega-\Omega\Lambda),
\end{align*}
where we have used again (\ref{potential}), which gives the desired
conclusion.
\end{proof}
Therefore, the value of
\[
\partial_n\Omega-L\Omega+\Omega\Lambda
\]
is a constant that does not vary with the KP flows. In particular,
choosing adequate initial conditions, we could take it as zero; i. e.,
\begin{equation}\label{constraint}
 -\res(\D^{-1}b\otimes\tilde{\cal L}\beta\D^{-1})=L\Omega-\Omega\Lambda.
\end{equation}

Now, is easy to prove
\begin{Th}
The vectorial Darboux transformation (\ref{hatu}) of a solution of the
$n$-th GD hierarchy defined by $b$ and $\beta$ as in (\ref{lax}) and
$\Omega$ satisfying (\ref{constraint}) gives again a solution of the
$n$-th GD hierarchy.
\end{Th}
\begin{proof}
As we have already commented we only need to check that the $f$'s are
independent of $t_n$.  This  follows from
\begin{align*}
\D_n(f_j)=&\langle \D_n\beta, \Omega^{-1} b^{(j)}\rangle
-\langle \beta, \Omega^{-1}(\D_n\Omega)\Omega^{-1} b^{(j)}\rangle
+\langle \beta,\Omega^{-1}(\D_nb^{(j)})\rangle\\
=&-\langle \beta, \Omega^{-1}\big(\Omega\Lambda +\D_n\Omega-
 L\Omega\big)\Omega^{-1}b^{(j)}\rangle\\
=&0.
\end{align*}
\end{proof}

\section{Reductions to the KPI and KPII hierarchies}

There are two main real reductions of the KP equation known as the
KPI and KPII equations, both with physical relevance. We show
here how the vectorial Darboux scheme presented above reduces to
these two real reductions. First, let us introduce the KPI and KPII
hierarchies:

\begin{itemize}
  \item The KPI hierarchy appears when
  \[
  t_{2n}\in\I\R,\;\; t_{2n-1}\in\R,\;\; n\in\N ,
   \]
and $u_{2,0}(\bt)\in\R$ takes real values. From the classical
results on the KP hierarchy one concludes that this
is equivalent to the following identity:
\[
(B_n)^*=(-1)^{n}\tilde B_n ,
\]
between the complex conjugate of the Zakharov-Shabat operators and
its  adjoint.

  \item The KPII case corresponds to real dependent and independent variables
  $t_n\in\R$, $n\in \N$, and $u_{2,0}$ real, that means real $B$'s and real
  $\tilde B$'s.
\end{itemize}

As for the GD reduction, the idea here is to
choose $b$, $\beta$ and
$\Omega$ in an appropriate way.

\subsection{KPI hierarchy}
For  the KPI case we assume that our linear space $V$ is a Hilbert space so 
that
it has an adjoint operation $^\dagger$.
 The adjoint linear system for
$\beta$ can be written
\[
(\D_n)^*\beta=B_n^*(\beta),
\]
where we have used that $(\D_n)^*=(-1)^{n+1}\D_n$. So, if $b$ solves
the direct problem then $b^\dagger H$ solves the adjoint one, where
$H$ is any linear operator in $V$. This motivates us to choose
$\beta$ precisely in this form, and then we have for
$\Omega$:
\begin{align*}
H^\dagger\D_n\Omega&=-H^\dagger\res(\D^{-1}b\otimes\tilde B_n
b^\dagger
\D^{-1})H,\\
\D_n\Omega^\dagger H&=-H^\dagger\res(\D^{-1}b \otimes (-1)^{n}B_n^* 
b^\dagger
\D^{-1})H\\&=-H^\dagger\res(\D^{-1}b \otimes \tilde B_n b^\dagger
\D^{-1})H,
\end{align*}
and so we find the relation
\[
\D_n(H^\dagger\Omega-\Omega^\dagger H)=0,\;\;\; n\in\N.
\]
With this we consider  the imaginary part  of $f_0$:
\[
f_0-f_0^*=\langle b^\dagger, (\Omega^\dagger)^{-1}(\Omega^\dagger
H-H^\dagger\Omega)\Omega^{-1}b\rangle,
\]
so if the constant operator $\Omega^\dagger H-H^\dagger\Omega$ is
chosen as zero; i. e., take the integration constant $\Omega_0$ for
$\Omega$ such that $\Omega_0^\dagger H=H^\dagger\Omega_0$, the
function
$f_0$ is real and therefore so is $\hat u_{2,0}$ and the vectorial
Darboux transformation preserves the KPI hierarchy reduction.

This can be resumed in the following:

\begin{pro}\label{kp1}
Let $B_n$ be a solution of the KPI hierarchy, $b\in V$ be a
vectorial solution of
\[
\D_n b=B_n(b),
\]
$H$ be a linear operator and $\Omega\in\GL(V)$ be an invertible
 linear operator defined by the
compatible equations
\[
\D_n\Omega=(-1)^{n+1}\res(\D^{-1}b \otimes B_n^* b^\dagger\D^{-1}) H,
\]
with initial condition $\Omega_0$ such that
\[
\Omega_0^\dagger H=H^\dagger\Omega_0,
\]
then the new $\hat B_n$ also solves the KPI hierarchy.
\end{pro}

Apart from the reality condition just shown, we are also interested
in those  vectorial Darboux transformations that preserve the
singularity structure of the solution; i. e., $\det\Omega\neq 0$. From
now on we take $H=\id$ so that $\Omega$ must be Hermitian; i. e.,
$\Omega^\dagger=\Omega$, and $\dim V<\infty$.

\begin{pro}\label{>}
Let $b$ be as in Proposition \ref{kp1} with its components be
linearly independent and such that the improper integral
\[
\int_{-\infty}^x\dif x\, b(x)\otimes b^\dagger(x),
\]
converges for all $x\in\R$. Then $\Omega$ can be taken as
\[
\Omega(x)=C+\int_{-\infty}^x\dif x\, b(x)\otimes b^\dagger(x),
\]
where  $C$ is a  hermitian constant operator. Moreover, if $C$ is
positive semi-definite, we have
\[
\det\Omega>0.
\]
\end{pro}

\begin{proof}
By construction $\Omega(x)=C+\int_{-\infty}^x\dif x\, b(x)\otimes
b(x)^\dagger$, where the constant hermitian operator $C$ is
positive semi-definite (all its eigen-values non-negative). Now,
the matrix $\int_{-\infty}^x\dif x\, b(x)\otimes b(x)^\dagger$ is
precisely a Gramm matrix for functions in the interval
$(-\infty,x]$. An elementary result from linear algebra
\cite{gelfand} tells us that it must be positive semi-definite  and
the determinant vanishes only if the functions $\{b_1,\dotsc,b_N\}$
are linearly dependent.

\end{proof}

Let us remark that for zero-background similar solutions can be
found in \cite{nm}, in fact there the solutions to the heat hierarchy
are expressed in integral form. However, there is no proof on the absence
of singularities.

\paragraph{Examples: line solitons, standard and exotic lumps.}
For the zero background $u_{2,0}=0$, that is $B_n=\D^n$, the linear
system for $b$ is just the free heat hierarchy: $\D_mb=\D^m(b)$. An
obvious solution is $\exp(\sum_{m\geq 1} t_mk^m)$ where $k\in\C$ is
an arbitrary complex number. Polynomial solutions can be expressed
as finite linear combination of the Schur or heat polynomials
$s_n(\bt)$ which can be constructed from
\[
\exp(\sum_{n\geq 1} t_nk^n)=\sum_{m\geq 0} s_m(\bt)k^m,
\]
or equivalently form the recursion relation
\[
s_n(\bt)=\frac{1}{n}\sum_{j=1}^nj t_j s_{n-j}(\bt), \;\;\; s_0(\bt)=1,
\]
being the first few:
\begin{align*}
  s_0(\bt) &= 1, \\
  s_1(\bt) &= t_1,\\
  s_2(\bt) &= t_2+\frac{1}{2}t_1^2,\\
  s_3(\bt) &= t_3+t_2t_1+\frac{1}{6}t_1^3,\\
  s_4(\bt) &= t_4+t_3t_1+\frac{1}{2} t_2(t_2+t_1^2)+\frac{1}{24}t_1^4.
\end{align*}
We recall the Galilean transformation: if $b(\bt)$ is a solution so
is
\[
b(\tilde{\bt}(k))\exp(\sum_{n\geq 1} t_n k^n),\;\;\; \tilde
t_n(k):=\sum_{m\geq n} \binom{m}{n}k^{m-n}t_m.
\]
Thus, performing a Galilean transformation on the heat polynomials
one gets  {\em heat  quasipolynomials}.

We assume $V$ to be $N$-dimensional and take $b=(b_1,\dotsc,b_N)^t$
with $b_j(\bt)=p_j(\tilde{\bt}(k_j))\exp(\sum_{n\geq 1}k_j^n t_n)$
a heat quasipolynomial, that is $p_j(\bt)$ is a heat polynomial of
order $n_j$; here we take distinct $k_j$ with $\re k_j>0$. Observe
that this choice for $b$ fulfils the requirements of Proposition
\ref{>} and therefore the associated solution will be non-singular.
As a result we get the following family of non-singular solutions
of the KPI hierarchy:
\begin{pro}
Let $\Omega$ be  the following hermitian matrix
\[
\Omega=C+{\cal E}\omega{\cal E}^\dagger,
\]
with $C$ a non-negative definite hermitian matrix and
\[
{\cal E}:=\operatorname{diag}(\exp(\sum_{n\geq 1} t_n
k_1^n),\dotsc,\exp(\sum_{n\geq 1} t_n k_N^n)),
\]
 and
\[
\omega_{ij}(\bt)=\frac{1}{k_i+k_j^*}
\sum_{\ell=0}^{n_i+n_j}\Big(-\frac{1}{k_i+k_j^*}\D\Big)^\ell
(p_i(\tilde{\bt}(k_i))p_j^*(\tilde{\bt}(k_j)),
\]
where $p_j$ is a heat polynomial of degree $n_j$ and
 $\tilde t_n(k)=\sum_{m\geq n} \binom{m}{n}k^{m-n}t_m$.
 Then,
 \[
 u=2\D^2\ln\det\Omega,
 \]
is a non-singular solution of the KPI hierarchy. When $C=0$ we get
the following non-singular rational solution
\[
u=2\D^2\ln\det\omega.
\]
\end{pro}

The $N$-line soliton solution of KPI equation appears once we take
the simplest heat polynomial; i. e., $p_i=a_i$ a complex constant
and $\omega_{ij}=a_ia_j^*/(k_i+k_j^*)$, and $C=\id$. If one allows
higher heat polynomials we get non-singular rational deformation of
the line solitons whenever $C\neq 0$. See \cite{naka,miya}.

However, when $C=0$ the situation changes drastically. Then
 we obtain non-singular rational
solutions of the KPI equation. In fact, for $N=1$ and $n_1=1$ we
get the 1-lump solution \cite{mz,as,sa} for KPI and for $n_1=2$ the
Johnson-Thompson solution \cite{jt} recently studied in \cite{av}.
For $N=1$ we can increase the degree $n_1$ of the heat polynomial
$p_1$, as is already suggested \cite{jt}, and get more involved
rational behaviour showing, as is studied in \cite{av}, non-trivial
interaction of localized lumps. For $N>1$ the $\tau$-function
$\det\omega$ could be understood as the non-linear composition of
$N$ of these objects.

For the 1-lump solution \cite{mz,as} we  choose $b$ as
\[
b=(x+2iky+3k^2t) \exp(kx+ik^2y+k^3t),
\]
with $k=k_{\text{R}}+\I k_{\text{I}}\in\C$, $k_{\text{R}},k_{\text{I}}\in\R$
by direct calculation, we have the following $\tau$-function
\[
 \omega=
\frac{1}{2k_{\text{R}}}\bigg[ \big(z(t)-\frac{1}{2k_{\text{R}}}\big)^2+
k_{\text{R}}^2 y(t)^2+\frac{1}{4k_{\text{R}}^2}
\bigg],
\]
with $x(t):=x+3(k_{\text{R}}^2+k_{\text{I}}^2)t$,
$y(t):=y+3k_{\text{I}}t$ and $z(t):=x(t)-2k_{\text{I}}y(t)$.

For the Johnson-Thompson solution \cite{jt} we take $b$ as
\[
b=\big[(x+2iky+3k^2t)^2+2(\I y+3k t)\big] \exp(kx+ik^2y+k^3t),
\]
again by direct calculation, we obtain
\begin{align*}
 \omega=&
\frac{1}{2k_{\text{R}}}\Bigg[ \bigg\{\Big(z(t)-
\frac{1}{2k_{\text{R}}}\Big)^2+\frac{1}{4k_{\text{R}}^2}+
6k_{\text{R}}t-4k_{\text{R}}^2y(t)^2\bigg\}^2\\ &
+16 k_{\text{R}}^2 y(t)^2z(t)^2+
\frac{1}{k_{\text{R}}^2}\Big(z(t)-
\frac{1}{2k_{\text{R}}}\Big)^2 +4y(t)^2+\frac{1}{4k_{\text{R}}^2}
\Bigg].
\end{align*}
After a translation in the $x$-coordinate this goes to the
solution presented in \cite{jt}.

Given a polynomial $p$ with degree $\deg p$, for negative $\alpha$
we have the identity
\[
\int_{-\infty}^x \dif x\, |p(x)|^2 \exp(-x/\alpha)=-\alpha \exp(-x/\alpha)
\sum_{m=0}^{\deg(p)}\alpha^{2m}\Bigg|\sum_{j=0}^{\deg(p)-m} \binom{j+m}{m}
\alpha^j p^{(j+m)}\Bigg|^2,
\]
that can be checked by integrating by parts the integral and using
the Leibnitz rule and observing that the result coincides with the
expansion of the right hand side term. With this at hand, and the
previous proposition for $\dim V=1$ we deduce an explicit formula
for  the function $\omega$ where its positivity is explicitly
shown, and thus the non-singularity property is explicit:

\begin{pro}
Given a heat polynomial $p$ and $k\in\C$, with $\re k>0$, the
function
\[
\sum_{m=0}^{\deg(p)}\frac{1}{(2\re k)^{2m}}\Bigg|\sum_{j=0}^{\deg(p)-m} 
\binom{j+m}{m}
(-1)^j\frac{1}{(2\re k)^j} p^{(j+m)}(\tilde{\bt}(k))\Bigg|^2,
\]
is  a $\tau$-function for the KPI hierarchy that gives rise to a
rationally non-singular localized solution of the KPI hierarchy.
\end{pro}

In particular, if $p=s_n$ is a Schur polynomial of order $n$ we get
the following $\tau$-function:
\[
\sum_{m=0}^{n}\frac{1}{(2\re k)^{2m}}\Bigg|\sum_{j=0}^{n-m} \binom{j+m}{m}
(-1)^j\frac{1}{(2\re k)^j} s_{n-j-m}(\tilde{\bt}(k))\Bigg|^2,
\]
for $n=1,2$ we recover the 1-lump and the Johnson-Thompson
solution.

%

\subsection{KPII hierarchy}

For the KPII case the procedure is quite different, now the reality
of the $B_n$ and $t_n$ ensures that if $b$ is a solution so is
$Pb^*$, where $P$ is a linear operator, and the same for $\beta $
and $\beta ^*Q$. As before we could say that
\begin{align*}
b^*&=Pb,\\
\beta&=\beta^*Q,
\end{align*}
but this is consistent only when $P$ and $Q$ satisfy
 $P P^*=Q Q^*=\id$. Let us suppose that this
is the case and analyze $\Omega$:
 \begin{align*}
  P\D_n\Omega &= -P\res(\D^{-1}b\otimes \tilde B_n\beta^*\D^{-1})Q, \\
  \D_n\Omega^*Q &= -P\res(\D^{-1}b\otimes \tilde B_n \beta^*\D^{-1})Q,
  \end{align*}
and therefore
\[
\D_n(P\Omega-\Omega^*Q)=0,\;\; n\in\N.
\]
Then, if the integration constant $\Omega_0$ satisfies
$P\Omega_0=\Omega_0^*Q$ so does $\Omega$. If we make this choice
then we have
\[
f_0-f_0^*=\langle
\beta^*,{\Omega^*}^{-1}(\Omega^*Q-P\Omega)\Omega^{-1}b\rangle=0,
\]
so $f_0$ is real, and so is the new solution. Thus, we conclude
that with this choice the KPII hierarchy is preserved by the
vectorial Darboux transformation.

\begin{pro}
Let $B_n$ be a solution of the KPII hierarchy, and $b\in V$ be a
vectorial solution of
\[
\D_n b=B_n(b),
\]
such that $b^*=Pb$ and $\beta\in V^*$ be  solution of the adjoint
system
\[
\D_n \beta=-\tilde B_n(\beta),
\]
such that $\beta=\beta^*Q$, where $P,Q\in\GL(V)$ with $P P^*= Q
Q^*=\id$,
 and let
$\Omega\in\GL(V)$ be a linear operator defined by the compatible
equations
\[
\D_n\Omega=-\res(\D^{-1}b\otimes \tilde B_n \beta\D^{-1}),
\]
with initial condition $\Omega_0$ such that
\[
\Omega_0^* Q=P\Omega_0,
\]
then the new $\hat B_n$ also solves the KPII hierarchy.
\end{pro}

When $P=Q=\id$ we have real  $b$, $\beta$ and $\Omega$. The line
soliton solution of the KPII hierarchy appears when the entries of
the $b$ and of the $\beta$ are just exponentials. Another
possibility is to take $V=W\oplus W$, even dimensional space and
take
\[
P=Q=\begin{pmatrix} 0&\id_W\\ \id_W&0\end{pmatrix},
\]
with respect to the just written splitting of $V$. This means that
\begin{align*}
b&=\begin{pmatrix}g\\g^*\end{pmatrix},\\
\beta&=(\gamma,\gamma^*),
\end{align*}
 where $g(\bt)\in W$ and $\gamma(\bt)\in W^*$, and
\[
\Omega=\begin{pmatrix}
{\cal A}&{\cal B}\\ {\cal B}^* &{\cal A}^*\end{pmatrix},
\]
with
\begin{align*}
\D_n{\cal A}&=-\res(\D^{-1}g\otimes \tilde B_n\gamma\D^{-1}),\\
\D_n{\cal B}&=-\res(\D^{-1}g\otimes \tilde B_n\gamma^*\D^{-1}).
\end{align*}
For example, for $W=\C$ we get
\[
\det\Omega=|{\cal A}|^2-|{\cal B}|^2.
\]
Taking $g(\bt)=p(\tilde{\bt}(k))\exp(\sum_n k^n t_n)$ and
$\gamma(\bt)=q(\tilde{\bt}(l))\exp(-\sum_n l^n t_n)$ where $p,q$
are heat polynomials, with degrees $\deg(p),\deg(q)$ and $k,l\in
\C$ with $\re k,\re l>0$, we have
\begin{align*}
  {\cal A} &=A+\exp\left(\sum_n (k^n-l^n) t_n\right)\sum_{m=0}^{\deg(p)+
  \deg(q)}\frac{(-1)^m}{(k-l)^{m+1}}(pq)^{(m)},\\
  {\cal B} &=B+\exp\left(\sum_n (k^n-(l^*)^n) t_n\right)\sum_{m=0}^{\deg(p)+
  \deg(q)}\frac{(-1)^m}{(k-l^*)^{m+1}}(pq^*)^{(m)},
\end{align*}
where $A,B$ are arbitrary complex constants. If these two constants
do not vanish we have a singular solution of mixed rational and
exponential character. However, as for KPI, when $A=B=0$ we have a
rational solution
\[
\det\Omega(\bt)=\exp\big(\sum_n 2\re(k^n-l^n) t_n\big)\tau(\bt),
\]
with the $\tau$-function given by
\[
\tau:=
\bigg|\sum_{m=0}^{\deg(p)+
  \deg(q)}\frac{(-1)^m}{(k-l)^{m+1}}(pq)^{(m)}\bigg|^2-\bigg|\sum_{m=0}^{\deg(p)+
  \deg(q)}\frac{(-1)^m}{(k-l^*)^{m+1}}(pq^*)^{(m)}\bigg|^2.
\]
The singularities lay over algebraic curves in the plane defined by
${\cal A} = \mu{\cal B}$, with $|\mu|=1$. For the KPII equation
these algebraic curves move in the plane, $x=t_1, y=x_2$, as the
time $t=t_3$ evolves. Similar results for the
Davey-Stewartson II are given in \cite{app}. Let us remind that the
motion of these singularities in $y$ is given by a Calogero-Moser (CM)
system while the one in $t$ is a higher order CM system, this constitutes the
Krichever theorem, see \cite{nm}.

\section{Reductions to Gel'fand-Dickey I and II}

We now discuss these reductions of type I and II in relation with
the GD reduction, that is two real reductions of the GD
hierarchy, which we denote by GDI and GDII. Observe that the KPI and
KPII reductions are compatible with the GD reduction and that for
the KdV case we have that the two reductions coincide. The question
here is whether  there is a vectorial Darboux transformation for
these GDI and GDII hierarchies. In order to get such result we
just need to combine in an adequate manner the reductions schemes
explained above.

First we analyze the GDI hierarchy. On the one hand the $n$-th GD
reduction requests $b$ and $\beta$ to satisfy the equations ${\cal
L}(b)=Lb$ and $\tilde{{\cal L}}(\beta)=-\beta\Lambda$, where ${\cal
L}=B_n$ and $L,\Lambda\in\text{L}(V)$ are spectral operators. On
the other hand the KPI reduction asks $b$ and $\beta$ to satisfy
$\beta=b^\dagger H$, $H$ an arbitrary linear operator in $V$. The
point is to make these two requirements compatible, and this can be
done if the operators $L,\Lambda$ and $H$ are linked by
\[
(-1)^nL^\dagger H+H\Lambda=0,
\]
that when $H$ is invertible means that
$\Lambda=(-1)^nH^{-1}L^\dagger H$, and so $\Lambda$ is expressed in
terms of $L,H$. If this is so, then ${\cal L}(b)=Lb$ implies
$\tilde{{\cal L}}(\beta)=(-1)^n({\cal L}(b))^\dagger H=b^\dagger
(-1)^nL^\dagger H=-\beta \Lambda$. As we know, the fact that $b$
and $\beta=b^\dagger H$ satisfy the spectral equations ensures, that
for adequate initial  conditions, the following equation holds
\[
(-1)^{n+1}\res(\D^{-1}b\otimes{\cal
L}^*b^\dagger\D^{-1})H=L\Omega-\Omega
\Lambda,
\]
we also know, from the KPI reduction, that it is possible to take
initials conditions so that
\[
\Omega^\dagger H=H^\dagger\Omega,
\]
so if it is possible to find an initial condition $\Omega_0$ so
that
\begin{align*}
  (-1)^{n+1}\res(\D^{-1}b\otimes{\cal L}^*b^\dagger\D^{-1})_0H & =L\Omega_0-
\Omega_0
\Lambda, \\
  \Omega_0^\dagger H &= H^\dagger\Omega_0,
\end{align*}
we know that the vectorial Darboux transformation preserves the
$n$-th GDI hierarchy.
\begin{pro}
Let ${\cal L}$ be a solution of the $n$-th GDI hierarchy and
$b(\bt)\in V$ a solution of the linear system
\begin{align*}
  {\cal L}(b) &= Lb, \\
  \D_n b & =B_n(b),
\end{align*}
with $L$ a linear operator in $V$ and take linear operators
$\Lambda$ and $H$ subject to
\[
(-1)^nL^\dagger H+H\Lambda=0.
\]
Find an invertible operator by integrating the compatible equations
\[
\D_n\Omega=-\res(\D^{-1}b\otimes\tilde{B}_nb^\dagger\D^{-1})H,
\]
with an initial condition $\Omega_0$ so that
\begin{align*}
  (-1)^{n+1}\res(\D^{-1}b\otimes{\cal L}^*b^\dagger\D^{-1})_0H& =L\Omega_0-
\Omega_0
\Lambda, \\
  \Omega_0^\dagger H &= H^\dagger\Omega_0,
\end{align*}
then the associated vectorial Darboux transformation preserves the
$n$-th GDI hierarchy.
\end{pro}

For the GDII hierarchy one proceeds in a similar way. The
conditions on $P,Q,L$ and $\Lambda$ are
\begin{align*}
  L^*P &=PL, \\
  \Lambda^*Q &=Q\Lambda,
\end{align*}
that make the conditions ${\cal L}(b)=Lb$, $\tilde{{\cal
L}}(\beta)=-\beta\Lambda$ compatible with $b^*=Pb$,
$\beta=\beta^*Q$. So, the conclusion is:
\begin{pro}
Let ${\cal L}$ be a solution of the $n$-th GDII hierarchy,
$b(\bt)\in V$ a solution of the linear system
\begin{align*}
  {\cal L}(b) &= Lb, \\
  \D_n b & =B_n(b),
\end{align*}
and $\beta(\bt)\in V^*$ a solution of
\begin{align*}
  \tilde{{\cal L}}(\beta) &=- \beta\Lambda,\\
  \D_n \beta & =-\tilde B_n(b)
  \end{align*}
  with $L,\Lambda$ linear operators in
$V$ and take linear operators $P,Q$ subject to $PP^*=QQ^*=\id$ and
\begin{align*}
  L^*P &=P L, \\
  \Lambda^* Q &=Q\Lambda.
\end{align*}
Find an invertible operator by integrating the compatible equations
\[
\D_n\Omega=-\res(\D^{-1}b\otimes\tilde{B}_n\beta\D^{-1}),
\]
with an initial condition $\Omega_0$ so that
\begin{align*}
-\res(\D^{-1}b\otimes\tilde{{\cal L}}\beta\D^{-1})_0&
=L\Omega_0-\Omega_0
\Lambda, \\
\Omega_0^* Q&=P\Omega_0,
\end{align*}
then the associated vectorial Darboux transformation preserves the
$n$-th GDII hierarchy.
\end{pro}

While for the 1st GD hierarchy, the KdV hierarchy, both real
reductions coincide with the real KdV hierarchy, and so we have two
different vectorial Darboux transformations for the same hierarchy,
in general the GDI and GDII reductions are different. In
particular, the 2nd GD hierarchy is associated with ${\cal
L}=\D^3+u_1\D+u_0$, and the Boussinesq equation is linked with the
$t_2$ flow given by $B_2=\D^2+u_{2,0}$. The compatibility
condition: $\D_2{\cal L}=[B_2,{\cal L}]$ gives
 $u_{2,0}=:u$, $\D u_1=3/2\D u$, so that
we can take  $u_1=3/2 u+c$, $c$ an arbitrary constant and
$\D(u_0)=3/4(\D_2(u)+\D^2(u))$ and the following equation is
satisfied:
\[
3\D_2^2u+4c\D^2(u)+3 \D^2(u^2)+\D^4(u)=0.
\]
The Boussinesq I equation takes real $u$ and $x\in\R$ and $t_2=\I
t$, $t\in\R$, while the Boussinesq II is associated with real $u$
and $x,t_2=t\in\R$. The Boussinesq I and II equations are
\[
\pm 3\D_t^2u= 4c\D^2(u)+3 \D^2(u^2)+\D^4(u).
\]
with $+$ corresponding to type I and $-$ to type II. Both equations
have been considered in the literature, but only Boussinesq I
appears to have physical applications, see \cite{ac} and references therein.

\end{document}